\documentclass[prl,twocolumn,aps,showpacs]{revtex4}
\usepackage{graphicx}
\usepackage{amsmath}
\usepackage{amsfonts}
\def\be{\begin{equation}}
\def\ba{\begin{eqnarray}}

\def\b{\beta}

\def\d{\delta}

\def\r{\rho}

\def\f{\varphi}

\def\ps{\psi}

\def\i{\int}

\def\bq{{\mathbf q}}
\def\bp{{\mathbf p}}
\def\bz{{\mathbf z}}

\def\bp{{\mathbf p}}
\def\bq{{\mathbf q}}

\def\cT{{\mathcal T_>}}
\def\cH{{\mathcal H}}

\def\Tr{\mbox{Tr}}

\def\ee#1{\label{#1}\end{equation}}
\def\ea#1{\label{#1}\end{eqnarray}}

\begin{document}
\title{Fluctuation theorems: Work is not an observable }
\author{Peter Talkner, Eric Lutz, Peter H\"anggi}
\affiliation{Institute of Physics, University of Augsburg,
D-86135 Augsburg, Germany}
\date{\today}
\begin{abstract}
The characteristic function of the work performed by an
external time-dependent force on
a Hamiltonian quantum system is identified with the {\it time-ordered
  correlation function} of the exponentiated system's Hamiltonian. A similar
expression is obtained for the averaged exponential work which is
related to the free energy difference of equilibrium systems 
by the Jarzynski work theorem.
\end{abstract}
\maketitle

Deep relations between nonequilibrium fluctuations and thermal
equilibrium properties of small systems have recently been discovered and 
formulated
in terms of so-called fluctuation theorems \cite{Ja06,Ja07}. These theorems are
not only of basic theoretical relevance  but provide novel
ground 
for experimental investigations of small systems in physics,
chemistry and biology \cite{Bu05} 

In this rapid communication we want to preclude a possible confusion about the
notion of work in the context of fluctuation theorems for quantum mechanical
Hamiltonian systems. To be precise, we consider a quantum  system 
which was in
thermal contact with a heat bath at inverse temperature $\b$ until a time
$t_0$. Then the contact with the bath is switched off and a
classical force acts on the otherwise isolated 
Hamiltonian system until the time
$t_f$ according to a prescribed protocol. We demonstrate that the
exponential average of the total work performed 
on the system as
well as the characteristic function of this work are given by {\it time-ordered
correlation functions} of the exponentiated Hamiltonian rather than by
expectation values of an operator representing the work as a pretended
observable.     

For a system that evolves under the exclusive influence of a time
dependent Hamiltonian $H(t)$ from an initial thermal equilibrium state
\be
\r(0) = Z(0)^{-1} \exp \left \{ - \b H (0) \right \}, 
\ee{r0}
at time $t_0=0$ 
until a final time $t = t_f$, the work performed on the system is a
randomly distributed quantity $w$. Its statistical properties follow
from a probability density $p(w)$ or, equivalently, from  the
corresponding characteristic function $G(u)$ which is defined as the
Fourier transform of the probability density, i.e.
\be
G(u) = \i dw e^{i u w} p(w)
\ee{G}
We will demonstrate that the characteristic function is given by 
the following quantum correlation function:
\be
\begin{split}
G(u)&= \langle e^{i u H(t_f)} e^{-i u H(0)} \rangle  \\
&\equiv \Tr \:e^{i u H_H(t_f)}e^{ -i u H(0)} \r(0)  \\ 
&= \Tr \:e^{i u H(t_f)} U(t_f) e^{ -i u H(0)} \r(0) U^+(t_f)
 \\
&= \Tr\: e^{iu H(t_f)} U(t_f) e^{-(i u+\b) H(0)} U^+(t_f)/Z(0)
\end{split}
\ee{GC}
where $\Tr$ denotes the trace over the system's Hilbert space $\cH$,
$U(t)$ the unitary time evolution governed by the
Schr\"odinger equation $i \hbar \partial U(t)/\partial t = H(t) U(t)$ with
$U(0) = 1$, and $H_H(t) = U^+(t) H(t) U(t)$ is the
Hamiltonian in the Heisenberg picture. The third equality follows
from the second line by the cyclic invariance of the trace and the
last line follows with eq.~(\ref{r0}). 

The characteristic function has the form of a time-ordered correlation
function of the two operators $\exp \{i u H(t_f) \}$ and $\exp \{-i u
H(0) \}$. We note that this correlation function in general  {\it
  differs} from the
averaged exponential of the difference of the Hamiltonians 
$W = H_H(t_f) - H(0)$, which sometimes is referred to as the operator of
work \cite{AN05}. 
It is possible though to formally rewrite the characteristic function
in terms of the difference between the Hamiltonians. In the second
line of eq.(\ref{GC}) the product of the operators $\exp \{iu H_H(t) \}$ 
and  $\exp\{-iu H(0) \}$ occurs in chronological order and may be
written as $ \exp \{iu H_H(t) \}\exp \{-iu H(0) \} = \cT \exp \{iu
H_H(t) \}\exp \{-iu H(0) \}$. Under the protection of the time ordering
operator $\cT$ the usual rule for exponentials of commutative quantities
holds \cite{Z68}
to yield the following equivalent forms of the characteristic function of work
\ba
G(u) &=& \Tr\: \cT e^{iu (H_H(t_f) - H(0))} \r(0) \nonumber \\
&=& \Tr\: \cT \exp \{iu \i_0^{t_f} \frac{\partial H_H(s)}{\partial s} ds
\} \r(0)
\ea{GT}
The second equality is a consequence of the known fact that the total
derivative of the Hamiltonian in the Heisenberg picture coincides with
its partial time derivative.

The averaged exponentiated  work $\langle \exp \left \{ - \b w \right 
\} \rangle$ is obtained from the characteristic function by putting $u = i
\b$, cf. eq. (\ref{G}). Using the correlation function expression
(\ref{GC}) together 
with the canonical initial density matrix (\ref{r0}) we immediately
recover the Jarzynski equation in its known form \cite{JaPRL97}
\be
\langle e^{-\b w} \rangle = \frac{Z(t_f)}{Z(0)}
\ee{JE}
where $Z(t_f) = \Tr\: e^{-\b H(t_f)}$ is the partition function of a 
hypothetical system with Hamiltonian $H(t_f)$ in a Gibbs state at
inverse temperature $\b$. 

By replacing the quantum correlation function by the
corresponding correlation function of a classical Hamiltonian system
the characteristic function of the work
performed on the classical system is obtained. Its 
inverse Fourier transform yields 
the known classical expression 
for the probability density of work, cf. Ref.~\cite{JaPRE97}
 
\ba
p_{\text{ cl}}(w) &=& Z_{\text{cl}}^{-1}(0) \i d\bz(0) \nonumber \\
&&\d \big (w -
[H(\bz(t_f),t_f)-H(\bz(0),0)]\big ) \nonumber \\ 
&&\times e^{-\b H(\bz(0),0)}
\ea{rcl} 
where $Z_{\text{cl}}(0) = \i d\bz \exp \{ -\b H(\bz,0) \}$ denotes the
classical partition function, $\bz =(\bp,\bq)$ a point in phase space
which serves as the initial condition of the trajectory $\bz(t)$ evolving
according to Hamilton's equations of motion.

The fluctuation theorem has
long been known for a sudden switch of the Hamiltonian of a
classical system \cite{Z54}. 
For a quantum system with a Hamiltonian changing from $H_0$ at time $t_0=0^-$ to
$H_1$ at $t_f=0^+$  the time-evolution operator becomes $U(t_f) =
1$. The characteristic function~(\ref{GC}) then simplifies to
\be
G(u) = \Tr \: e^{i u H_1} e^{-iu H_0} e^{-\b H_0}/Z(0)
\ee{Gs}
and the Jarzynski equation~(\ref{JE}) again follows  
with $u=i\b$.
In all nontrivial cases, when the two Hamiltonians do not commute, 
the averaged exponential of the difference operator $H_1-H_0$ 
does not yield this result.

The proof of eq.~(\ref{GC}) essentially follows an argument given by Kurchan
\cite{K00}, see also Refs. \cite{Ta00,Mu03,Mo05}. It is based on the
elementary observation that
two energy measurements are required in order to determine
the work performed on the system by an external force.
In the first measurement, the energy is determined 
in the initial Gibbs state $Z(0)^{-1} \exp \{- \b
H(0) \}$. The outcome of this measurement is one of the eigenvalues $e_n(0)$ of
the Hamiltonian $H(0)$
with the probability 
\be
p_n = \exp \{ - \b e_n(0) \}/Z(0). 
\ee{pn}
After the measurement
the system is found in the corresponding eigenstate $\f_n(0)$ of
$H(0)$ satisfying $H(0) \f_n(0) = e_n(0) \f_n(0)$. This state
evolves according to $\ps(t) = U(t) \f_n(0)$ until the second 
energy measurement is performed at the time $t_f$. It produces an eigenvalue
$e_m(t_f)$ with the probability 
\be
p(m,t_f|n) = |\big (\f_m(t_f)|U(t_f)
  \f_n(0) \big )|^2, 
\ee{pnm}
where $(\cdot|\cdot )$ denotes the
  scalar product of the Hilbert space $\cH$. Here, $e_m(t_f)$ and
  $\f_m(t_f)$ are the eigenvalues and eigenfunctions, respectively, of
  the Hamiltonian $H(t_f)$. 
Hence, the energies
$e_m(t_f)$ and $e_n(0)$ are measured with the probability  $p(m,t|n) p_n$     
such that the probability density of the work, which is the difference
of the measured energies, becomes
\be
p(w) = \sum_{n,m} \d\big (w-[e_m(t_f)-e_n(0)] \big )\: p(m,t_f|n) p_n
\ee{rw}

One then finds for the characteristic function from the definition (\ref{G})
\be
\begin{split}
G(u) =& \sum_{n,m} e^{i u (e_m(t_f) - e_n(0))}  \\
&\times \big (\f_m(t_f) |U(t_f) \f_n(0) \big )  \\
&\times \big (\f_n(0)|U^+(t_f)  \f_m(t_f) \big ) e^{-\b e_n(0)}/Z(0)
\\ 
= &\sum_{n,m} \big ( \f_m(t_f) |U(t_f) e^{-i u H(0)} \r(0)\f_n(0) \big )
\\
&\times \big (\f_n(0)|U^+(t_f) e^{i u H(t_f)} \f_m(t_f) \big )
\\
=& \Tr \:U(t_f) e^{-iuH(0)} \r(0) U^+(t_f) e^{iuH(t_f)}
\end{split}
\ee{pGC}
In going to the last line, the sum over the complete set of
eigenstates $\{\f_m(t)\}$ was written as the trace and the completeness of
the eigenstates $\f_n(0)$ was used. By use of the cyclic invariance of the
trace the  
quantum correlation function expression (\ref{GC}) for the characteristic
function is proved.  

This expression for the characteristic function contains all available
statistical information about the work 
performed by an external force on an isolated quantum system, such as
the averaged exponentiated work, cf.~eq. (\ref{JE}). All moments of the
work follow from the derivatives of the characteristic function.  

{\it Acknowledgements} This work was supported by the SFB, project A10
(PH, PT),  
the Nanosystems Initiative Munich (NIM), and the Emmy Noether Program
of the DFG under contract LU1382/1-1 (EL).

\end{document}